\documentclass[12pt,twocolumn]{aastex63} 
\usepackage{amssymb}
\usepackage{amsmath}
\usepackage{amsmath}
\usepackage{hyperref}
\usepackage{graphics,graphicx,natbib}
\usepackage{cleveref}
\usepackage{enumitem} 
\usepackage{hhline} 

\begin{document}

\newcommand{\kwm}[1]{{\color{purple}#1}}
\newcommand{\kwmcomment}[1]{{\color{purple}[#1]}}

\title{An Injection System for the CHIME/FRB Experiment}
\shorttitle{}
\shortauthors{}

\author[0000-0003-2095-0380]{Marcus Merryfield}
  \affiliation{Department of Physics, McGill University, 3600 rue University, Montr\'eal, QC H3A 2T8, Canada}
  \affiliation{McGill Space Institute, McGill University, 3550 rue University, Montr\'eal, QC H3A 2A7, Canada}
\author[0000-0003-2548-2926]{Shriharsh P.~Tendulkar}
  \affiliation{Department of Astronomy and Astrophysics, Tata Institute of Fundamental Research, Mumbai, 400005, India}
  \affiliation{National Centre for Radio Astrophysics, Post Bag 3, Ganeshkhind, Pune, 411007, India}
\author[0000-0002-6823-2073]{Kaitlyn Shin}
  \affiliation{MIT Kavli Institute for Astrophysics and Space Research, Massachusetts Institute of Technology, 77 Massachusetts Ave, Cambridge, MA 02139, USA}
  \affiliation{Department of Physics, Massachusetts Institute of Technology, 77 Massachusetts Ave, Cambridge, MA 02139, USA}
\author[0000-0001-5908-3152]{Bridget Andersen}
  \affiliation{Department of Physics, McGill University, 3600 rue University, Montr\'eal, QC H3A 2T8, Canada}
  \affiliation{McGill Space Institute, McGill University, 3550 rue University, Montr\'eal, QC H3A 2A7, Canada}
\author[0000-0003-3059-6223]{Alexander Josephy}
  \affiliation{Department of Physics, McGill University, 3600 rue University, Montr\'eal, QC H3A    2T8, Canada}
  \affiliation{McGill Space Institute, McGill University, 3550 rue University, Montr\'eal, QC H3A   2A7, Canada}
\author[0000-0003-1884-348X]{Deborah Good}
  \affiliation{Department of Physics, University of Connecticut, 196 Auditorium Road, U-3046, Storrs, CT 06269-3046, USA}
  \affiliation{Center for Computational Astrophysics, Flatiron Institute, 162 5th Avenue, New York, NY 10010, USA}
\author[0000-0003-4098-5222]{Fengqiu Adam Dong}
  \affiliation{Department of Physics and Astronomy, University of British Columbia, 6224 Agricultural Road, Vancouver, BC V6T 1Z1 Canada}
\author[0000-0002-4279-6946]{Kiyoshi W. Masui}
  \affiliation{MIT Kavli Institute for Astrophysics and Space Research, Massachusetts Institute of Technology, 77 Massachusetts Ave, Cambridge, MA 02139, USA}
  \affiliation{Department of Physics, Massachusetts Institute of Technology, 77 Massachusetts Ave, Cambridge, MA 02139, USA}
\author[0000-0002-1172-0754]{Dustin Lang}
  \affiliation{Perimeter Institute of Theoretical Physics, 31 Caroline Street North, Waterloo, ON N2L 2Y5, Canada}
  \affiliation{Department of Physics and Astronomy, University of Waterloo, Waterloo, ON N2L 3G1, Canada}
\author[0000-0002-3777-7791]{Moritz M\"{u}nchmeyer}
  \affiliation{Department of Physics, University of Wisconsin-Madison, 1150 University Ave, Madison, WI 53706, USA}
\author[0000-0002-1800-8233]{Charanjot Brar}
  \affiliation{Department of Physics, McGill University, 3600 rue University, Montr\'eal, QC H3A 2T8, Canada}
  \affiliation{McGill Space Institute, McGill University, 3550 rue University, Montr\'eal, QC H3A 2A7, Canada}
\author[0000-0003-2047-5276]{Tomas Cassanelli}
  \affiliation{Department of Electrical Engineering, Universidad de Chile, Av. Tupper 2007, Santiago 8370451, Chile             }
\author[0000-0001-7166-6422]{Matt Dobbs}
  \affiliation{Department of Physics, McGill University, 3600 rue University, Montr\'eal, QC H3A 2T8, Canada}
  \affiliation{McGill Space Institute, McGill University, 3550 rue University, Montr\'eal, QC H3A 2A7, Canada}
\author[0000-0001-8384-5049]{Emmanuel Fonseca}
  \affiliation{Department of Physics and Astronomy, West Virginia University, PO Box 6315, Morgantown, WV 26506, USA }
  \affiliation{Center for Gravitational Waves and Cosmology, West Virginia University, Chestnut Ridge Research Building, Morgantown, WV 26505, USA}
\author[0000-0001-9345-0307]{Victoria M.~Kaspi}
  \affiliation{Department of Physics, McGill University, 3600 rue University, Montr\'eal, QC H3A 2T8, Canada}
  \affiliation{McGill Space Institute, McGill University, 3550 rue University, Montr\'eal, QC H3A 2A7, Canada}
\author[0000-0002-0772-9326]{Juan Mena-Parra}
  \affiliation{MIT Kavli Institute for Astrophysics and Space Research, Massachusetts Institute of Technology, 77 Massachusetts Ave, Cambridge, MA 02139, USA}
\author[0000-0002-4795-697X]{Ziggy Pleunis}
  \affiliation{Dunlap Institute for Astronomy \& Astrophysics, University of Toronto, 50 St.~George Street, Toronto, ON M5S 3H4, Canada}
\author[0000-0001-7694-6650]{Masoud Rafiei-Ravandi}
  \affiliation{Department of Physics, McGill University, 3600 rue University, Montr\'eal, QC H3A 2T8, Canada}
  \affiliation{McGill Space Institute, McGill University, 3550 rue University, Montr\'eal, QC H3A 2A7, Canada}
\author[0000-0003-3154-3676]{Ketan R.~Sand}
  \affiliation{Department of Physics, McGill University, 3600 rue University, Montr\'eal, QC H3A 2T8, Canada}
  \affiliation{McGill Space Institute, McGill University, 3550 rue University, Montr\'eal, QC H3A 2A7, Canada}
\author[0000-0002-7374-7119]{Paul Scholz}
  \affiliation{Dunlap Institute for Astronomy \& Astrophysics, University of Toronto, 50 St.~George Street, Toronto, ON M5S 3H4, Canada}
\author[0000-0002-2088-3125]{Kendrick Smith}
  \affiliation{Perimeter Institute of Theoretical Physics, 31 Caroline Street North, Waterloo, ON N2L 2Y5, Canada}
\author[0000-0001-9784-8670]{Ingrid H.~Stairs}
  \affiliation{Department of Physics and Astronomy, University of British Columbia, 6224 Agricultural Road, Vancouver, BC V6T 1Z1 Canada}
\newcommand{\allacks}{
F.A.D is supported by the four year fellowship
FRB research at UBC is supported by an NSERC Discovery Grant and by the Canadian Institute for Advanced Research.
J.M.P is a Kavli Fellow.
K.S. is supported by the NSF Graduate Research Fellowship Program.
K.W.M. is supported by an NSF Grant (2008031).
M.D. is supported by a Canada Research Chair, Killam Fellowship, NSERC Discovery Grant, CIFAR, and by the FRQNT Centre de Recherche en Astrophysique du Qu\'ebec (CRAQ)
M.Me. is supported by an NSERC PGS-D award.
M.Mu. is supported by a DOE grant (DE-SC0022342)
P.S. is a Dunlap Fellow.
V.M.K. holds the Lorne Trottier Chair in Astrophysics \& Cosmology, a Distinguished James McGill Professorship, and receives support from an NSERC Discovery grant (RGPIN 228738-13), from an R. Howard Webster Foundation Fellowship from CIFAR, and from the FRQNT CRAQ.
Z.P. is a Dunlap Fellow.
}

\correspondingauthor{Marcus Merryfield}
\email{marcus.merryfield@mail.mcgill.ca}

\begin{abstract}
    \centering Dedicated surveys searching for Fast Radio Bursts (FRBs) are subject to selection effects which bias the observed population of events. Software injection systems are one method of correcting for these biases by injecting a mock population of synthetic FRBs directly into the realtime search pipeline. The injected population may then be used to map intrinsic burst properties onto an expected signal-to-noise ratio (SNR), so long as telescope characteristics such as the beam model and calibration factors are properly accounted for. This paper presents an injection system developed for the Canadian Hydrogen Intensity Mapping Experiment Fast Radio Burst project (CHIME/FRB). The system was tested to ensure high detection efficiency, and the pulse calibration method was verified. Using an injection population of $\sim$85,000 synthetic FRBs, we found that the correlation between fluence and SNR for injected FRBs was consistent with that of CHIME/FRB detections in the first CHIME/FRB catalog. We also noted that the sensitivity of the telescope varied strongly as a function of the broadened burst width, but not as a function of the dispersion measure. We conclude that some of the machine-learning based Radio Frequency Interference (RFI) mitigation methods used by CHIME/FRB can be re-trained using injection data to increase sensitivity to wide events, and that planned upgrades to the presented injection system will allow for determining a more accurate CHIME/FRB selection function in the near future.
\end{abstract}

\section{Introduction}

Fast radio bursts, or FRBs, were first discovered through searching archival data from a pulsar survey using the 64-m Parkes Murriyang Radio Telescope \citep{lbm+07}. This first detection, dubbed the ``Lorimer burst,'' bore characteristics of pulses emitted by pulsars, but with a dispersion measure (DM) indicating it must have originated far outside the Galaxy. Since the discovery of the Lorimer burst, the extragalactic nature of FRBs has been confirmed \citep{clw+17, tbc+17} and the field of FRB science has bloomed \citep[see][for a recent review of the field]{phl21}. Many radio telescopes have now detected FRBs and several dedicated FRB surveys are currently underway.

A dedicated FRB survey from the Canadian Hydrogen Intensity Mapping Experiment \citep[CHIME,][]{CHIME_overview22, chime_overview_2018}, has led to a recent abundance of detected FRBs. The first catalog from the CHIME Fast Radio Burst project \citep[CHIME/FRB,][]{chime_catalog_1} contains a total of 536 bursts from 492 sources, and more are constantly being discovered. CHIME/FRB’s design allows for the highest FRB detection rate of all current surveys but, like any telescope, it is subject to selection biases (some of which vary with time or with RFI evolution) affecting the observed population of bursts.

One way to correct for biases introduced in astrophysical transient searches is for telescopes to develop systems which inject synthetic events into their detection pipelines. By injecting pulses over a broad parameter space and tracking their (non-)detection status, it is possible to construct a probability distribution of event detection. In principle this selection function can be determined by injecting a sufficiently large number of simulated events to cover the entirety of the observable, multi-dimensional parameter space. With a large enough sample, the selection function would be given by the detection fraction across fine-grain histogram bins spanning the full volume. In the case of FRBs, thoughtful approximations must be made as the FRB parameter space is very large. Injection systems also have other uses besides determining the selection function. For example, injected events test the efficiency and robustness of complex data analysis systems, and provide a method of continuous system sensitivity monitoring.

Though there are examples of injection systems elsewhere in astronomy \citep[such as in the Laser Interferometer Gravitational-Wave Observatory, see][]{ligo_injections17}, injection systems for FRB searches are not yet in common use. Currently, an injection system exists for the Molonglo Observatory Synthesis Telescope and has been used to quantify biases in their FRB search pipeline \citep{vcw+21}. An injection analysis has also been performed at Arecibo for the Pulsar Arecibo L-band Feed Array survey single pulse pipeline \citep{pab+18}. Other such pipelines are in development at the Very Large Array \citep{lbb+18} and at the Green Bank Telescope \citep{als+20}. However, the first CHIME/FRB catalog is currently the only FRB dataset which accounts for selection effects based from the results of an injection study \citep{chime_catalog_1}. Specifically, an injection system was developed which injected $\sim$85,000 synthetic FRBs into the CHIME/FRB detection pipeline. Input parameters for each synthetic pulse were stored, as were output parameters for detections successfully matched to injections. These injections demonstrated, among other things, only modest selection effects with dispersion measure in the survey, but much larger selection effects with scattering time and intrinsic burst width.

In this paper, we present the synthetic pulse injection system used by CHIME/FRB in the injection study for the first CHIME/FRB catalog. We also provide the full dataset used to generate the CHIME/FRB selection function. An overview of the CHIME/FRB detection pipeline is provided in \S \ref{sec:chime}, and \S \ref{sec:injection_architecture} provides the details of the injection system architecture. An overview of how synthetic FRBs are modelled is given in \S \ref{sec:modelling}, and \S \ref{sec:injections} describes the input distributions used in our synthetic FRB population. Finally, in \S \ref{sec:results}, we note the various ways the injection system has been verified, including tests to ensure the synthetic FRBs are correctly flux calibrated. We also assess the efficiency of the pipeline, and note where the telescope is most susceptible to missing FRB detections.

\begin{table*}[!t]
    \centering
    \begin{tabular}{c c c c c}
        \hline
        Tree Index & DS factor & Widths searched (ms) & Max DM (pc cm$^{-3}$) & $\Delta$DM (pc cm$^{-3}$) \\ \hline
        0 & 1 & $1\sim3 \times 0.98304$ & 1656.13 & 1.62 \\
        1 & 2 & $1\sim3 \times 1.96608$ & 3312.27 & 3.23 \\
        2 & 4 & $1\sim3 \times 3.93216$ & 6624.55 & 6.47 \\
        3 & 8 & $1\sim3 \times 7.86432$ & 13249.11 & 12.94 \\
        4 & 16 & $1\sim3 \times 15.72864$ & 13249.11 & 25.88 \\
        \hline
    \end{tabular}
    \caption{Dedispersion trees searched by \texttt{bonsai}. For each downsampling tree, the downsampling factor, search widths, maximum searched DM, and width of the trees' DM bins ($\Delta$DM) are given. Note the downsampling factors, the temporal search kernels, the width of the DM search bins, and the maximum DM of each tree are configurable.}
    \label{tab:bonsai_pars}
\end{table*}

\section{The CHIME Telescope} \label{sec:chime}

The Canadian Hydrogen Intensity Mapping Experiment \citep[CHIME, see][]{CHIME_overview22} is a recently commissioned radio telescope near Penticton, British Columbia. CHIME was originally conceived solely as a cosmology experiment, with the goal of detecting a baryon acoustic oscillation (BAO) signal from neutral hydrogen at $z=0.8$--$2.5$, corresponding to observing frequencies from 400--800\ \text{MHz} \citep{chime_21cm_detection22}. The telescope design requires a large field of view ($\sim$200 deg$^2$) and sufficient sensitivity to detect the BAO signal. CHIME achieves this through its four 20-m $\times$ 100-m cylindrical reflectors, which have a combined collecting area of $\sim$8000 m$^2$. Each cylinder is outfitted with 256 dual polarization feeds suspended above their axes, for a total of 1,024 feeds. CHIME's powerful FX correlator uses the inputs from each feed to form 256 beams N-S via Fast Fourier Transform (FFT) beamforming and 4 beams E-W through exact phasing, for a total of 1,024 closely spaced stationary beams tiling the telescope's primary beam \citep[see][for details]{ng+17}. CHIME/FRB's beamforming pipeline output is sampled with 16,384 (hereafter 16k) frequency channels at 0.98304 ms cadence \citep[see][for more details]{chime_overview_2018}.

\subsection{The CHIME/FRB Detection Pipeline} \label{sec:chime_frb}
CHIME/FRB searches for FRBs using 132 computing nodes to form a pipeline which is divided into four ``Levels,'' L1--L4. L1 applies 128 of these nodes for RFI mitigation and CHIME/FRB's tree dedispersion search algorithm (\texttt{bonsai}). L2/L3 uses two more nodes for multi-beam grouping, further RFI mitigation, extragalactic event determination, and action determination algorithms. L4 uses the final two nodes for action implementation and hosting of databases and web interfaces. A detailed overview of the CHIME/FRB detection pipeline is available in \cite{chime_overview_2018}, whereas aspects relevant to this work are highlighted below.

\begin{figure*}[!t]
    \centering
    \includegraphics[width=\linewidth]{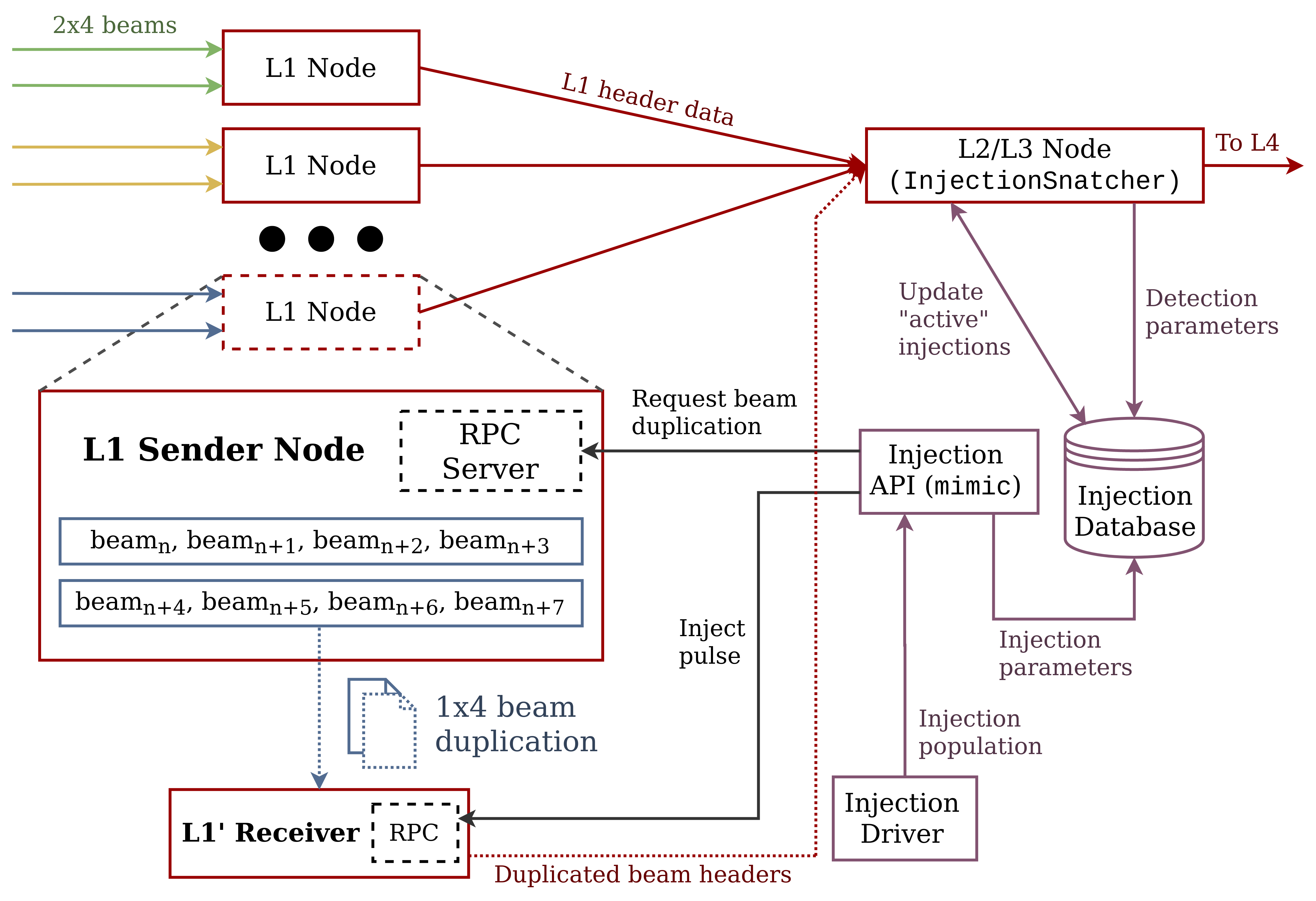}
    \caption{Schematic of the CHIME/FRB injection system. Each L1 node handles two datastreams containing full resolution intensity data for four beams. The injection API (\texttt{mimic}) interacts with the L1 nodes through Remote Procedure Calls (RPC). Requests to the API are made to send duplicated intensity data from four beams at a time to the L1$^\prime$ receiver. Once L1$^\prime$ has started receiving data, injections may be requested to any of the four beams. The detection properties of injections are then measured on L1$^\prime$, and the (non-)detection parameters are stored in a database alongside the input parameters.}
    \label{fig:injection_system}
\end{figure*}

\subsubsection{L1: RFI Mitigation and Dedispersion}

Each L1 node receives all 16k frequency channels for 8 beams. The intensity array is first sent through the initial (L1) RFI mitigation algorithm. This process involves both de-trending and ``clipping'' (truncation of extreme values, see Rafiei-Ravandi et al. 2022, in preparation). After initial RFI mitigation, \texttt{bonsai} searches for FRBs in intensity data, which is the most computationally expensive part of the L1 process. This involves tree dedispersion at the 0.98304 ms time resolution of CHIME/FRB and at four downsampled time resolutions, each binned by a factor of two up to a maximum temporal binning of $\sim$16 ms. For low DM pulses dedispersion is performed in all downsampling trees, where each ``tree'' is a 2D array of samples in $\left(\nu^{-2},\ t\right)$ space which represent pulses of various dispersion \citep[see][for more details]{tay74, 2012Hall}. Above the DM thresholds given in Table \ref{tab:bonsai_pars} only downsampled trees are searched. 

CHIME/FRB searches over a 4D parameter space (DM, $\beta$, $t$, $W$), where $\beta$, $t$, and $W$ are spectral index, arrival time of the pulse, and pulse width, respectively. DMs are searched to a maximum of $\sim$13000 pc cm$^{-3}$. Widths are searched using four search kernels with weights $[1]$, $[1,1]$, $[1,3,1]$, $[1,2,2,1]$, encompassing Full Width at Half Maximum (FWHM) widths up to approximately $3 \times \Delta t_{\text{tree}}$ (where $\Delta t_{tree}$ is the size of the time bin in the tree). Widths and DMs searched in each dedispersion tree are indicated in Table \ref{tab:bonsai_pars}. Two values of $\beta$ are searched, $\beta = \pm3$. Though a flat $\beta$ isn't searched, the chosen values of $\beta$ greatly improve SNR for pulses with emission located primarily in the top or bottom of the band, while maintaining near-optimal search conditions for flatter pulses.

The search output is binned substantially in DM and $t$ by taking the maximum SNR event within a bin to reduce the data volume. These ``coarse-grained'' events with SNRs greater than a tunable threshold are then vetted by a second RFI mitigation algorithm known as L1b. L1b discriminates between astrophysical signals and RFI using a neural network trained on a set of data human-labelled as astrophysical or RFI. A concise ``L1 header'' is then produced containing detection metadata, including the coarse-grained DM and $t$, an RFI rating from L1b, and SNR slices in DM and $\beta$ around the maximum SNR bin. Note that \texttt{bonsai} does not report which width kernel resulted in a trigger, but does report which tree the trigger was detected in. We thus report a nominal \texttt{bonsai} detection width of $2\times \Delta t_{\text{tree}}$ in our data.

\subsubsection{L2/L3: Grouping and Further RFI Mitigation}

After L1 processing, L1 headers for each beam are sent to L2/L3, which first attempts to group them in DM, $t$, and sky position to determine if a given event was detected in multiple beams. Events whose parameters all match within specified thresholds are grouped together. DM and time thresholds reflect the size of the coarse-grain L1 triggers, while spatial thresholds reflect the size of the CHIME/FRB beams. The grouped set of L1 headers becomes an L2 header, from which event positions are refined.

L2 then performs a second RFI excision based on the SNR distribution of L2 headers in neighboring beams. Astrophysical events, because they are in the far-field of the telescope, should have narrow SNR distributions within grouped beams. By contrast, RFI events are typically terrestrial, in the near-field, and have much broader SNR distributions within grouped beams. A machine learning classifier assigns an L2 RFI grade for each event. Aircraft transits are one major source of RFI, as they reflect terrestrial noise into the CHIME/FRB field of view and create RFI events across many beams for several minutes. Such events are very temporally wide and broadband. 

After RFI classification, the DM and sky location of the L2 events are compared to a database of known sources. If the event matches a source within a specified, tunable threshold, the event is labelled as associated with the known source. If there is no connection to a known source, the pulse's DM is compared to Galactic DM models \citep{ne2001,ymw17} to determine if the event is likely extragalactic. 

The last step in L2/L3 is to determine ``actions'' for each event based on criteria it fulfills. Most importantly, intensity and baseband data callbacks are queued for unknown extragalactic sources above a SNR threshold. These actions can (and have) been adjusted as improvements are made to the system. For example, the SNR threshold for intensity callback of unknown extragalactic candidates was $10\sigma$ in the early days of the experiment, $9\sigma$ for the majority the first CHIME/FRB catalog, and is currently $8\sigma$. 

\section{The Architecture of the Injection System} \label{sec:injection_architecture}

The CHIME/FRB injection system consists of multiple components: an application program interface (API) which sends injection requests to L1, an injections database which confers the status of injections with L2/L3, and an injection ``driver'' which manages the process of systematically injecting synthetic pulses. Note that rather than injecting simulated pulses into the live CHIME/FRB data stream, we make a copy of the data stream and then inject simulated pulses into the duplicated data stream. This avoids the possibility of injected data interfering with the live data stream.

The injection system is run as a Docker\footnote{\url{https://www.docker.com/}} stack on the CHIME/FRB analysis cluster. Docker is an OS-level virtualization service that is used to make CHIME/FRB container images, allowing for analysis tasks to be performed in a standardized runtime environment. This enables the CHIME/FRB analysis cluster, consisting of ten worker nodes, to easily run containerized tasks that otherwise may have had very different software dependencies. The injection system's API (\texttt{mimic}) is outlined in \S \ref{sec:mimic}.  Requests to inject synthetic pulses are sent to the API via the \texttt{InjectionDriver} (\S \ref{sec:driver}). The parameters of each requested pulse are stored in the Injection Database (\S \ref{sec:database}), which is queried in L2/L3 by the \texttt{InjectionSnatcher} (\S \ref{sec:snatcher}) in order to match incoming event headers with injections. The full set of interactions between the injection system and CHIME/FRB nodes is shown in Figure \ref{fig:injection_system}.

\subsection{The Injection API} \label{sec:mimic}

Injections are orchestrated by an API named \texttt{mimic}. This API is written in the \texttt{sanic}\footnote{\url{https://github.com/huge-success/sanic}} framework, and is naturally asynchronous, meaning it can handle many simultaneous requests from multiple clients. At its core, \texttt{mimic} manages injections in the realtime pipeline by interacting with L1 nodes via remote procedure calls (RPCs). A summary of how \texttt{mimic} interfaces with the realtime pipeline is given in Figure \ref{fig:injection_system}. The RPC servers outfitted on each L1 node have two crucial functions for the purposes of injections.

First, requests can be made to start (or stop) the duplication of on-sky data for one set of four intensity beams that are being processed by a given node. One of the 128 L1 nodes has been outfitted as a ``receiver node'' (L1$^\prime$) which processes duplicated beams from the ``sender node'' using the same software as the other L1 nodes. Though this reduces the amount of beams searched in the CHIME/FRB experiment (as the beams that would have been processed by L1$^\prime$ are replaced by duplicated beams), the L1 node processing the least sensitive beams in the array was chosen to host L1$^\prime$ so as to maintain as much exposure as possible.

Second, a request can be made to the L1$^\prime$ node to add a synthetic FRB into the datastream of one of the duplicated beams. The data from the receiver node are tagged with an arbitrary offset in their labelled beam numbers before being sent to L2/L3, which prevents the misclassification of injections as astrophysical signals. Because synthetic bursts are added into duplicated beams, CHIME/FRB remains nominally sensitive to astrophysical signals in the non-duplicated beams.

\subsection{The \texttt{InjectionDriver}} \label{sec:driver}

The \texttt{mimic} server only handles requests to start/stop beam duplication and inject pulses into duplicated beams. A separate module known as the \texttt{InjectionDriver} organizes a list of injection requests to be sent to the API. The \texttt{InjectionDriver} operates by reading an \texttt{hdf5}\footnote{\url{https://www.hdfgroup.org/solutions/hdf5/}} file containing a population of FRBs which have been generated for injection. The orchestration of these injections is more-or-less straightforward, but with some important technical considerations.

The outgoing L1 sender network can only handle duplicating one set of four beams at a given time, and \texttt{bonsai} requires a moderate amount of time ($\sim$several minutes) for its variance estimation of incoming data to reach a steady state. Given these two restrictions, the driver schedules injections in bundles of four beams. Each bundle's injections are run to completion before moving on to the next. This approach is not as ideal as fully randomizing the injections across all beams, because intermittent RFI could plague the detection statistics for injections in a given set of beams. However, it is time efficient since \texttt{bonsai}'s running variance estimation needs to stabilize each time a new set of beams is streamed to L1$^\prime$. Thus, going through four sequential sender beams at a time minimizes the waiting time.

The \texttt{InjectionDriver} handles four parallel threads at a time. Each thread is responsible for making requests to the injections API to generate pulses for one beam. The threads keep a running tally of the injections they have completed. The tallies from each thread are relayed to the \texttt{InjectionDriver}, which in turn keeps track of completed injections for a given schedule file. Such bookkeeping is important because networking issues or unplanned system shutdowns can otherwise interrupt injections.

\subsection{The Injection Database} \label{sec:database}


Injection and detection parameters for injections are stored in a \texttt{RethinkDB}\footnote{\url{https://rethinkdb.com/}} database hosted on the CHIME/FRB archiver. The database itself is not a part of the \texttt{mimic} stack, but rather is built as a part of another containerized project which manages various services and databases for CHIME/FRB. When injecting synthetic pulses, each injection is labelled with a unique ID. Further, a program name can be given to a set of injections to make the full results of any given run retrievable using the program name as a key.

The injection database's role is to store both the injection parameters and header detection parameters for each injection. A list of relevant parameters stored is given in Table \ref{tab:db_parameters}. \texttt{RethinkDB} is a good choice of architecture for the injections database because it is malleable -- if the capability of injections is expanded with new input parameters (e.g. scattering index $\alpha$), then previous database entries will remain unaffected. Within \texttt{RethinkDB} it is also possible to tag database entries with an expiration time. This is useful because it allows a list of active injections to be retrieved, which will expire if undetected by the \texttt{InjectionSnatcher} within the expiration time. After expiration, injections are removed from the list of active injections, and labelled as non-detections.
\begin{table}
    \centering
    \resizebox{\linewidth}{!}{
    \begin{tabular}{c c}
        \hline
        Injection Parameters & Units \\ \hline
        Expected Arrival Time & UTC timestamp \\
        Dispersion Measure & pc cm$^{-3}$ \\
        Fluence & Jy ms \\ 
        Width (Gaussian FWHM) & ms \\
        $\tau_\text{1 GHz}$ & ms \\
        Spectral index ($\beta$) & Unitless \\
        Spectral running ($\zeta$) & Unitless \\
        Beam number & Unitless \\
        Beam coordinates & Unitless \\ \hline \hline
        Detection Parameters & Units \\ \hline
        Detection Time & UTC Timestamp \\
        \texttt{bonsai} DM & pc cm$^{-3}$ \\
        \texttt{bonsai} SNR & Unitless \\
        \texttt{bonsai} downsampling tree & Unitless \\
        \texttt{bonsai} spectral index ($\beta^\prime$) & Unitless (May be $\pm 3$) \\
        L2 RFI Grade & Unitless (Graded from 0--10) \\ \hline
    \end{tabular}
    }
    \caption{A list of the injection parameters currently available as input to \texttt{mimic}, and relevant detection parameters for injections detected in the realtime pipeline.}
    \label{tab:db_parameters}
\end{table}

\subsection{The \texttt{InjectionSnatcher}} \label{sec:snatcher}


One of the most important aspects of the injection system is properly flagging injections as they come through the realtime pipeline, labelling them as either detected or not detected. The \texttt{InjectionSnatcher}, which acts as a module in the L2/L3 pipeline, fulfills this purpose. Built largely from the framework of the \texttt{KnownSourceSifter} \cite[see][]{chime_overview_2018, pleunis_thesis21}, the \texttt{InjectionSnatcher} inspects all incoming L2 event header metadata and compares their expected arrival times and DMs to the list of active injections in the database. If the arrival time of an L2 header matches within half a second of an active injection and if the DM of that same L2 header matches within twice the DM error of an active injection, then the detection is taken as a match to the injection. The detection parameters for the event are then sent to the database, matched with the corresponding injection parameters.

\begin{figure*}[!t]
    \centering
    \includegraphics[width=\linewidth]{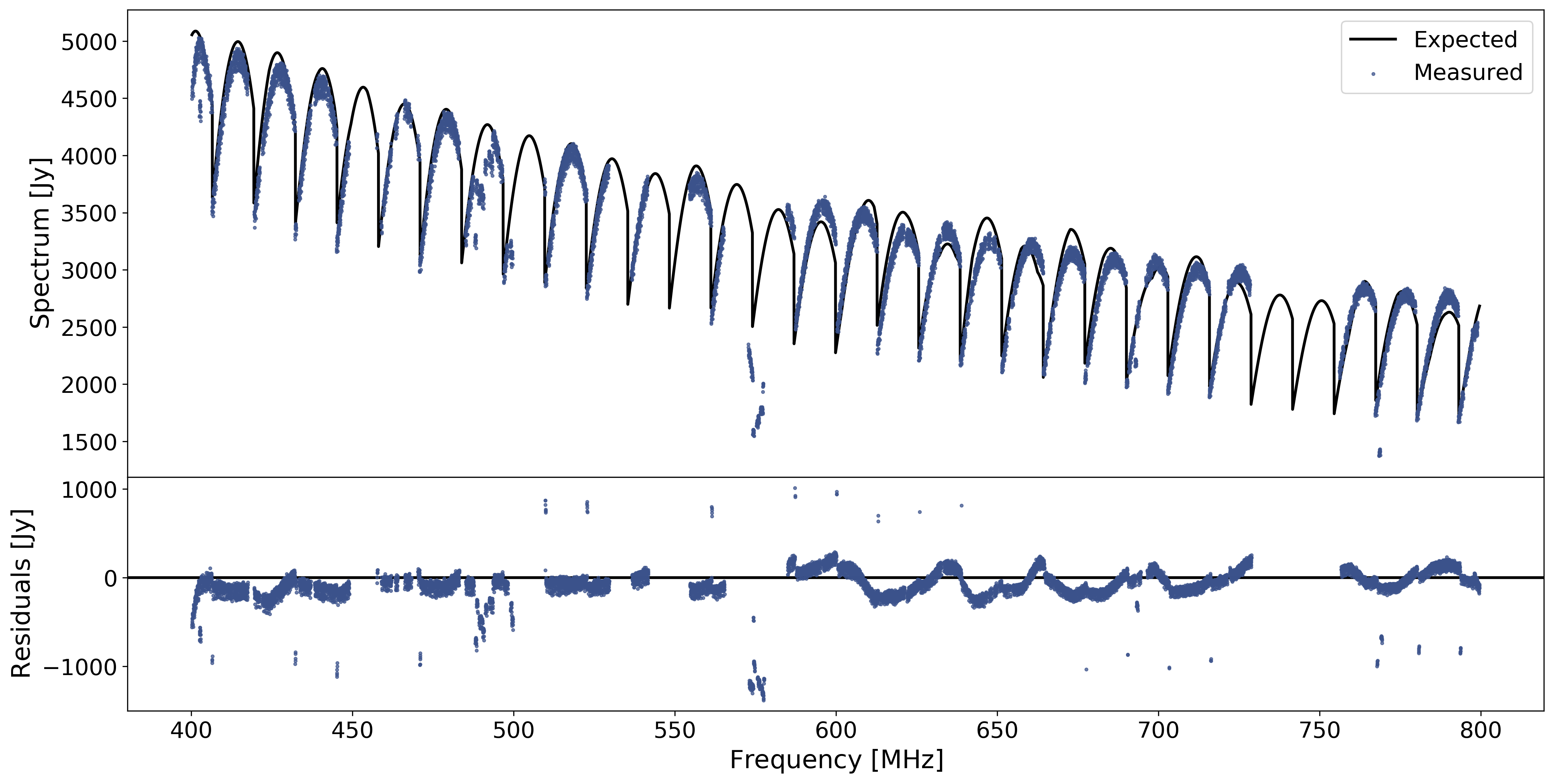}
    \caption{A comparison between the expected and measured fluence of Cygnus A from CHIME/FRB intensity data. The spectrum was observed on 2020 September 7, after the application of the $f_\text{good}^2$ factor in the gain conversion process. Frequencies without measured data points are excised due to contamination from RFI.}
    \label{fig:cyg_a_comparison}
\end{figure*}

\section{Modelling SYnthetic FRBs} \label{sec:modelling}

An important part of ensuring the injection system simulates realistic on-sky FRBs is the process of properly modelling both the FRB itself and the effects of the detection system. In the case of CHIME/FRB, injections are input into beamformed intensity data. Thus, beyond modelling the FRB itself, beamforming effects must be forward modelled.

\subsection{The \texttt{simpulse} Library} \label{sec:simpulse}


The \texttt{simpulse}\footnote{\url{https://github.com/kmsmith137/simpulse}} library generates dispersed, scattered FRBs, assuming Gaussian intrinsic burst widths. While the process of generating a dispersed pulse with a Gaussian width then convolving with a one-sided scattering exponential is simple, \texttt{simpulse} also forward models dispersion smearing, ensuring a realistic mock FRB. 

For now, \texttt{simpulse} assumes that the scattering index $\alpha$ is fixed to $-4$, and that pulse dispersion is exactly proportional to $\nu^{-2}$. Spectrally, FRB emission (particularly from repeating sources) is not always well described by a power law, as very band-limited patches of emission are often observed \cite[see e.g.][]{abb+19c}. This motivates instead the use of a ``running'' power law to describe FRB spectra, with an additional index $\zeta$ known as the spectral running \citep{p++21}:
\begin{equation} \label{eq:running_power_law}
    S_\nu \propto \left( \frac{\nu}{\nu_0} \right)^{\beta + \zeta \ln{(\nu / \nu_0)}},
\end{equation}
where $\nu_0$ is known as the ``pivotal frequency'' and is a chosen, fixed constant. 

\subsection{The Beam Model} \label{sec:beam_model}

Aside from an FRB's intrinsic properties, the injection system models the combined spectral signatures of the CHIME telescope's primary beam and the FFT synthesized beams. While the effect of the FFT synthesized beams is known precisely and can be calculated, this is not true of the telescope's primary beam. To model the telescope primary beam, the CHIME Intensity Mapping team developed a method which fits a coupled antenna model using a cylindrical reflector to holography data of steady source and solar transits. The results of the fit were saved to a file, which has been interpolated into a beam model module used for sensitivity calculations with CHIME/FRB \citep[see][for examples of the CHIME primary beam]{CHIME_overview22, CHIME_beams_sun}. We provide a public release of one of the early versions of the beam model used for the purposes of injections.\footnote{\url{https://github.com/chime-frb-open-data/chime-frb-beam-model}}

The beam model uses an $(x, y)$ coordinate system, which is topocentric. In these coordinates, $x$ and $y$ represent the projection of the sky on a unit circle drawn on the local ground plane, with zenith as the origin. So, an $(x,y)$ coordinate can be seen as a transformation of $(\text{RA}, \text{Dec})$ for a specific date and time. CHIME references its calibration to the position of CygA in the beam at meridian, $(x_\text{CygA}, y_\text{CygA})$, meaning the beam sensitivities for each of the 16k frequencies is fixed at unity at that location. From there, the sensitivity arrays are in units relative to the sensitivity at that location, with frequency averaged sensitivities at the center of each of the 1,024 beams ranging from $\sim$0.05$-$1.1 depending on the beam's location. The full array of sensitivities from the beam model are passed as input to the \texttt{simpulse} workers, and modulate the pulse before it is sent to duplicated beams for injection. The integrated fluence of the pulse is not conserved in this process, since the beam sensitivity affects the observed brightness.

\begin{figure*}[!t]
    \centering
    \includegraphics[width=0.9\linewidth]{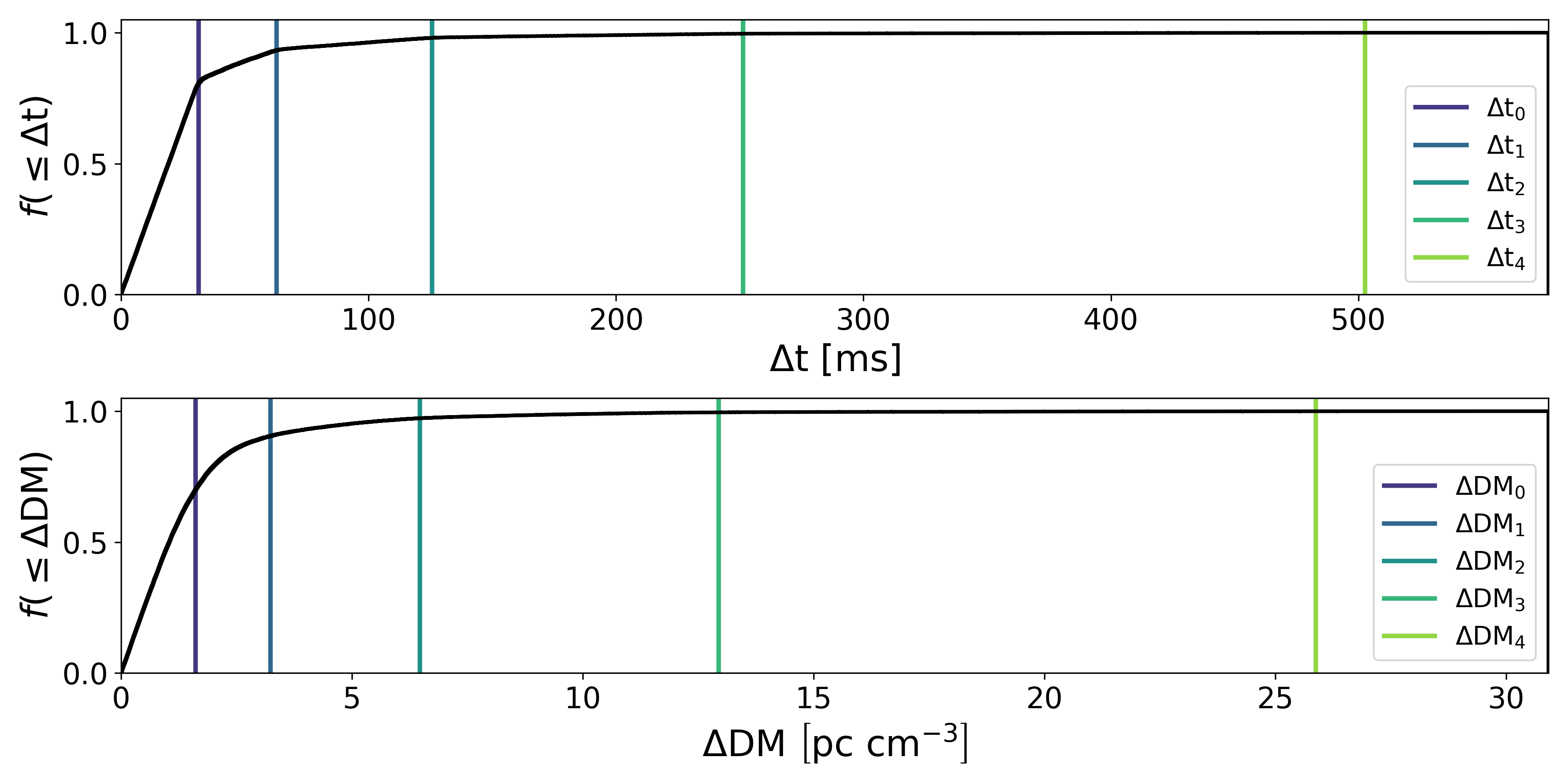}
    \caption{\textbf{Top:} A cumulative histogram showing the fraction of detected injections with respect to the absolute difference between the expected and detected arrival time of the injection ($f(\leq \Delta t),\ \Delta t = \left|t_i - t_d\right|$ where $t_i$ refers to the expected arrival time and $t_d$ refers to the detected arrival time). Colored lines show the expected detection time uncertainty for each L1 dedispersion tree, introduced as a result of the DM uncertainty from 400--800 MHz (0$-$4, see Table \ref{tab:bonsai_pars}). \textbf{Bottom:} A similar cumulative histogram for the absolute difference between the injected and detected DM of the injection ($\Delta \text{DM} = \left|\text{DM}_i - \text{DM}_d\right|$). Colored lines show the DM uncertainty for each L1 dedispersion tree (0--4, see Table \ref{tab:bonsai_pars}).}
    \label{fig:snatcher_verification}
\end{figure*}

\subsection{Calibration} \label{sec:calibration}

Ensuring that synthetic pulses are being injected in meaningful, physical energy units is a critical step in the process of simulating injections. Because CHIME calibrates daily on steady sources, beamformed intensity data for CHIME/FRB are calibrated up to a constant multiplied by the beam model. The total conversion factor is:
\begin{equation} \label{eq:cal_factor}
    C = \frac{(1024 f_{\text{good}})^2 \times 128}{4^2 \times 0.806745 \times 400}\ \ \text{BF units}\ \ \text{Jy}^{-1},
\end{equation}
where BF units are the units of beamformed CHIME/FRB data. Each of the numerical factors in Equation \ref{eq:cal_factor} comes from careful bookkeeping in the CHIME calibration process and the intensity data channelization:
\begin{itemize}
    \item The factor of $(1024 f_{\text{good}})^2$ represents the coherent sum of 1,024 antenna signals weighted by the number of CHIME feeds that are labelled as ``good'' during the array calibration process ($f_{\text{good}}$, typically ranges from 0.9 to 1.0). Feeds not labelled as good have an issue in their analog system, resulting in either a lack of signal or excess noise, and are given zero weight in the beamformer.
    \item The factor of 128 represents the incoherent sum of 128 time samples in the up-channelization process.
    \item The factor of $\frac{1}{4^2}$ comes from a factor of 4 introduced in the baseband data during the East-West beamforming process across CHIME's 4 cylinders.
    \item $\frac{1}{0.806745}$ is a factor that is introduced in the polyphase filter bank (PFB) bandpass correction \citep[see][]{price16}. The up-channelization process, coupled with the sample-to-sample correlations induced by the PFB, results in a factor-of-2 ripple in the bandpass on 400\,kHz scales, which must be corrected. The correction introduces the quoted overall scaling factor.
    \item The factor of $\frac{1}{400}$ is a configurable, deterministic factor that was chosen solely to adjust the scale of the downstream data. Its value has been static since CHIME/FRB's commissioning.
\end{itemize}

Given the calibration factor $C$, data can be converted between CHIME/FRB BF units and Jy units. When CHIME/FRB began its pre-commissioning phase, the input data from L0 to L1 did not account for the factor $f_\text{good}^2$. Since the value $f_\text{good}$ varies with time, the L1 intensity data could not be calibrated to a static factor in real time. However, in 2020 April, a correction for the factor $f_\text{good}^2$ was applied to the script which manages gain conversions, making CHIME/FRB intensity data calibrated in real time up to a static factor and beam model. The use of this calibration constant was tested against CygA transit data and compared to CygA's expected Jy spectrum, as shown in Figure \ref{fig:cyg_a_comparison}. The calibrated flux was consistently within $\sim$5\% of the expectation. 

\section{A Population of Synthetic FRBs} \label{sec:injections}

When generating a population of FRBs to be injected, a set of input FRB property distributions must be chosen. Ideally, the distribution of synthetic burst parameters should result in a reasonable fraction of detected events while still spanning many interesting parts of parameter space, such as large DM and/or scattering. Injection distributions should in principle be based on physical models, as then the model parameters in the true population of bursts could be better understood when comparing the results of injections to the detected FRB population. However, there are not yet physically well-motivated property distributions on which to model FRB populations.

\begin{figure}[t]
    \includegraphics[width=\linewidth]{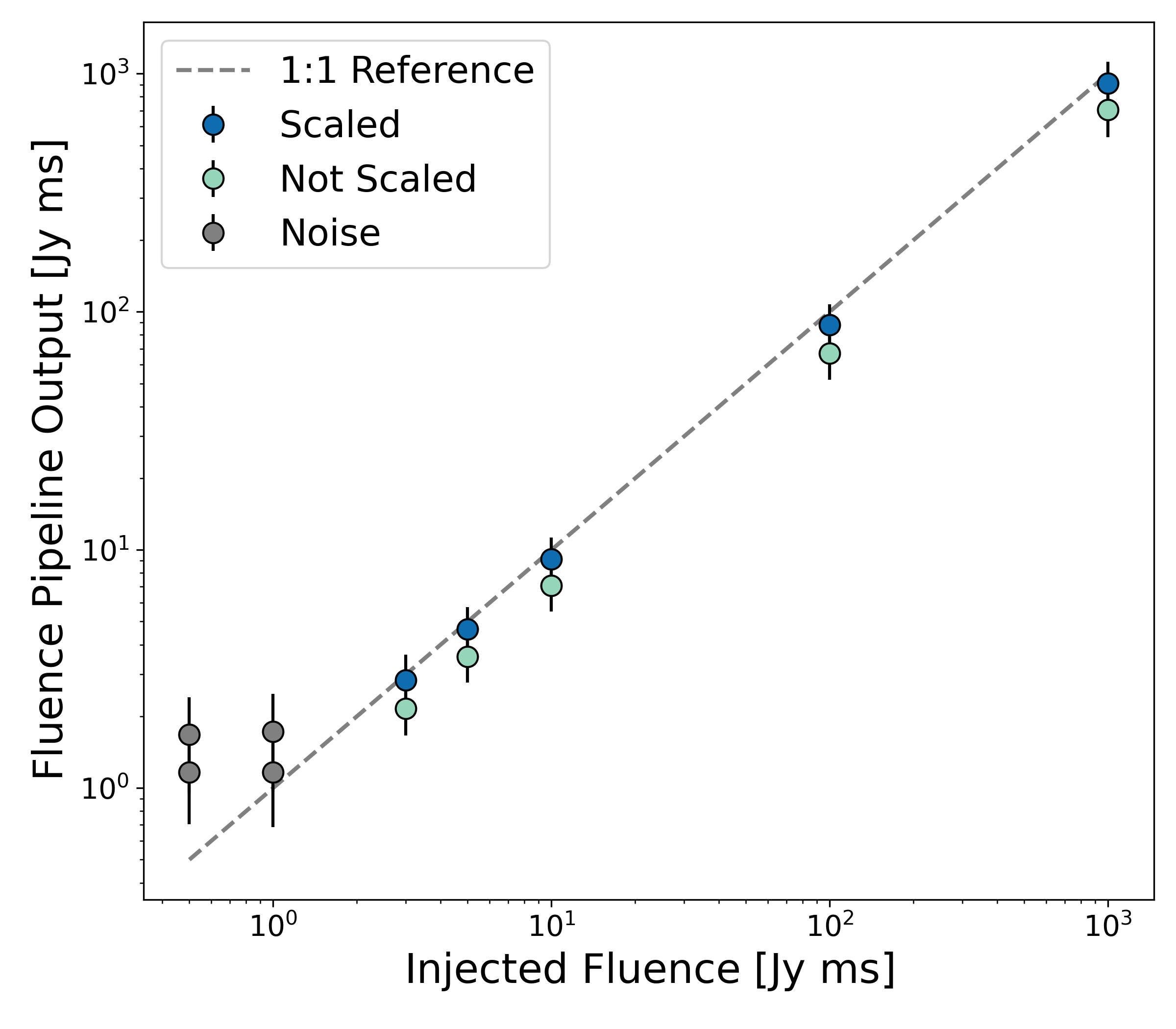}
    \caption{A comparison of the injected fluence and fluence pipeline output for a synthetic test pulse across a range of injected fluences. There are two sets of points for each injected fluence: the lighter blue points show the calibration pipeline output without correcting for the beam model (BM) sensitivities, whereas the darker blue points include a BM correction. Error bars represent an approximately 1$\sigma$ uncertainty. The dashed gray line shows where the fluence pipeline output would perfectly match the injection input fluence and the gray points show where the fluence pipeline began reporting the fluence of a noise peak instead of the synthetic pulse.}
    \label{fig:fluence_comp_test_event}
\end{figure}

In order to ensure synthetic bursts occupy an interesting region of parameter space while remaining detectable, CHIME/FRB currently uses kernel density estimation alongside a combination of uniform, log-normal, and power-law distributions, parts of which come from fits to the detection distributions in the first CHIME/FRB catalog (``catalog distributions''). The details of the distributions for the relevant injection parameters in Table \ref{tab:db_parameters} are described here briefly:
\begin{itemize}
    \item For the DM distribution, 90\% of the population is drawn from a log-normal fit to the catalog distribution, while 10\% of the population is drawn from a uniform distribution between 0 and 5000 pc cm$^{-3}$.
    \item For the fluence distribution, the population is drawn from a power-law distribution with index $\gamma=-1.0$. The power law distribution was chosen to be shallower than the Euclidean $\gamma=-1.5$ so as to more efficiently sample rare, bright bursts.
    \item For the intrinsic width distribution, 90\% of the population is drawn from a log-normal fit to the catalog distribution, while 10\% of the population is drawn from a uniform distribution between 0 and 100 ms.
    \item The distributions for the spectral indices ($\beta$ and $\zeta$) are drawn from a kernel density estimator fits to the catalog distributions.
    \item For the scattering distribution, 90\% of the population is drawn from a log-normal fit to the catalog distribution, while 10\% of the population is drawn from a uniform distribution between 0 and 100 ms.
\end{itemize}

For their spatial distribution, we randomly sample $10^6$ locations uniformly distributed on the celestial sphere in telescope coordinates (coordinates in which the telescope is fixed). Each coordinate is sampled independently. Of these, we discard all locations: that are below the horizon; for which the band-averaged primary beam response is less than $10^{-2}$ in beam model units; and for which the band-averaged response does not reach $10^{-3}$ in any of the 1,024 synthesized beams. The fraction of sky locations surviving these cuts is $f_{sky} = 0.0277$ (an important factor in the all-sky completeness accounting).

\begin{figure}[t]
    \centering
    \includegraphics[width=\linewidth]{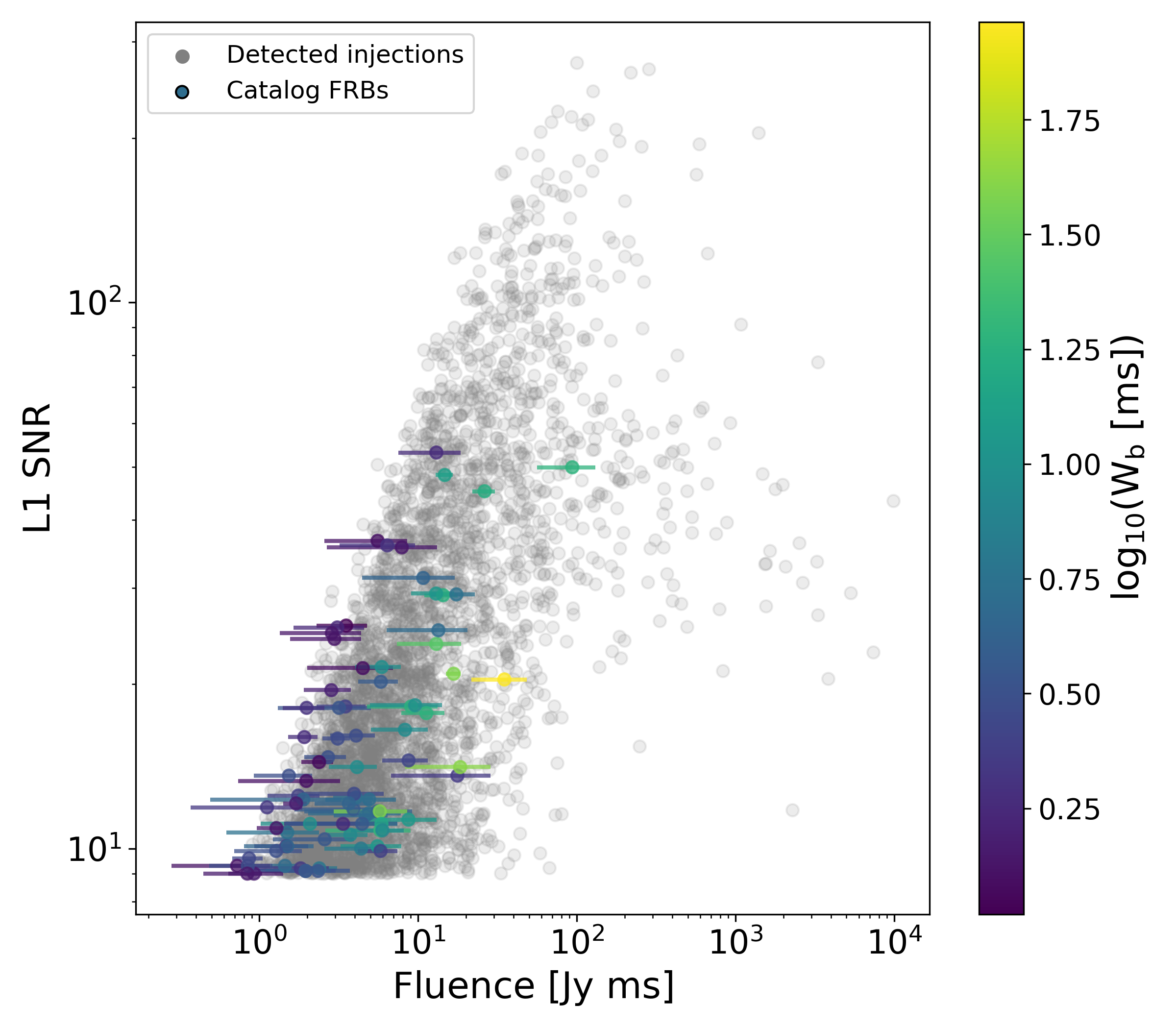}
    \caption{The fluence and SNR relationship for bursts in the first CHIME/FRB catalog (coloured by broadened width) and detected injections (in gray). Note a cut has been made to include only events within the FWHM sensitivity of the most sensitive beam column. This is because FRBs detected there have the most trustworthy fluences in the catalog sample.}
    \label{fig:fluence_v_snr}
\end{figure}

For the first CHIME/FRB catalog, a set of 5 million synthetic FRBs were first generated from the above distributions and randomly assigned to the surviving sky locations. For each FRB we calculate an upper limit on the SNR assuming the only radiometer noise is a system equivalent flux density (SEFD) of 45 Jy (on meridian at the declination of Cyg A). This is on the lower end of measurements of the SEFD, which vary from 30 to 80 Jy across the band. We account for the frequency-dependent beam model, FRB spectrum, and FRB temporal width, and assume an optimal signal-to-noise weighted average over frequency. In early trials we injected events for which this upper limit exceeded 9$\sigma$ (CHIME/FRB's detection threshold for the majority of the first catalog). However, the idealized assumptions in the limit turned out to be too conservative, resulting in a very low detection efficiency. As such, we raised this threshold to 20$\sigma$, resulting in an acceptable efficiency.

The result was a population whose sky distribution traced the beam sensitivity. This allowed not only for injections in the primary lobe of the CHIME/FRB beam, but also for very bright injections in the sidelobes.

Of the 5 million synthetic FRBs, $\sim$97,000 were determined to be potentially detectable by CHIME/FRB. The \texttt{InjectionDriver} requested the injection of these events between 2020 August 14 and 2020 September 22. Of the $\sim$97,000 events, only $\sim$85,000 were actually injected, as some injection requests were dropped due to networking issues. Of the successful injections, $\sim$30,000 were detected by L1 with SNR $>$ 9 and an L2 RFI grade $>$ 5 (the threshold set to distinguish RFI and astrophysical events), or an SNR $>$ 100 (a high SNR override put in place to avoid losing very bright events). The full dataset of 5 million synthetic FRBs, the subset of those that were injected, and a data usage tutorial is available at the CHIME/FRB Open Data portal.\footnote{\url{https://chime-frb-open-data.github.io/}}

\section{Results} \label{sec:results}


\subsection{Verifying the \texttt{InjectionSnatcher}}

Because the L1 system generates so many triggers, it was important to ensure the \texttt{InjectionSnatcher} was properly classifying injected bursts and not triggers from RFI. During the construction of the injection system, the \texttt{InjectionSnatcher} was tested for a set of several thousand simple, bright bursts to ensure its efficacy. The snatcher successfully retrieved these bursts with nearly 100\% detection efficiency, assuring its ability to match active injections with the corresponding detections.

A cumulative histogram of the absolute DM and time difference between injections and detections in the final injection sample is shown in Figure \ref{fig:snatcher_verification}. The large majority of events are detected within the first few trees' DM and time search widths, which gives confidence that few (if any) headers produced by RFI are being incorrectly classified as injections.

\subsection{Verifying Pulse Calibration}

Although the injection constant shown in Equation \ref{eq:cal_factor} was tested against steady source transit data, the calibration of synthetic pulses also needed to be verified. Since injections were performed after the L1 intensity data buffer, it was not possible to request callback data of injected FRBs and test their calibration using either CHIME/FRB's calibration pipeline or the calibration constant. Instead, calibration was tested offline by injecting a synthetic pulse into the intensity callback data of an FRB detected in one of the most sensitive CHIME/FRB beams after the implementation of the $f_\text{good}$ scaling.

The test pulse used to verify calibration was narrow ($\sim$1 ms) and unscattered, and injected across a range of fluences without dispersion, as FRB detections are dedispersed before being sent to the calibration pipeline. Sensitivities from the beam model were applied assuming the pulse was detected in the center of the formed beam. The resulting synthetic data were run through the CHIME/FRB calibration pipeline in order to compare the injection fluence and the pipeline output fluence \citep[see][for details on the CHIME/FRB fluence pipeline]{andersen_thesis21}. The comparison is shown in Figure \ref{fig:fluence_comp_test_event}. Normally the fluence estimation from the calibration pipeline does not take into account the primary beam model, which explains why the raw fluence pipeline output is systematically lower than the input fluence of the synthetic pulses. When a correction incorporating the sensitivity of the beam model is applied, the fluence pipeline output matches almost exactly the input fluence for most pulses, as expected. The two test pulses at and below the $1\ \text{Jy}\ \text{ms}$ input fluence level in Figure \ref{fig:fluence_comp_test_event} do not have reasonable fluence pipeline output as upon inspection those bursts were too faint, causing the pipeline to measure a noise peak.

\begin{figure}[t]
    \centering
    \includegraphics[width=\linewidth]{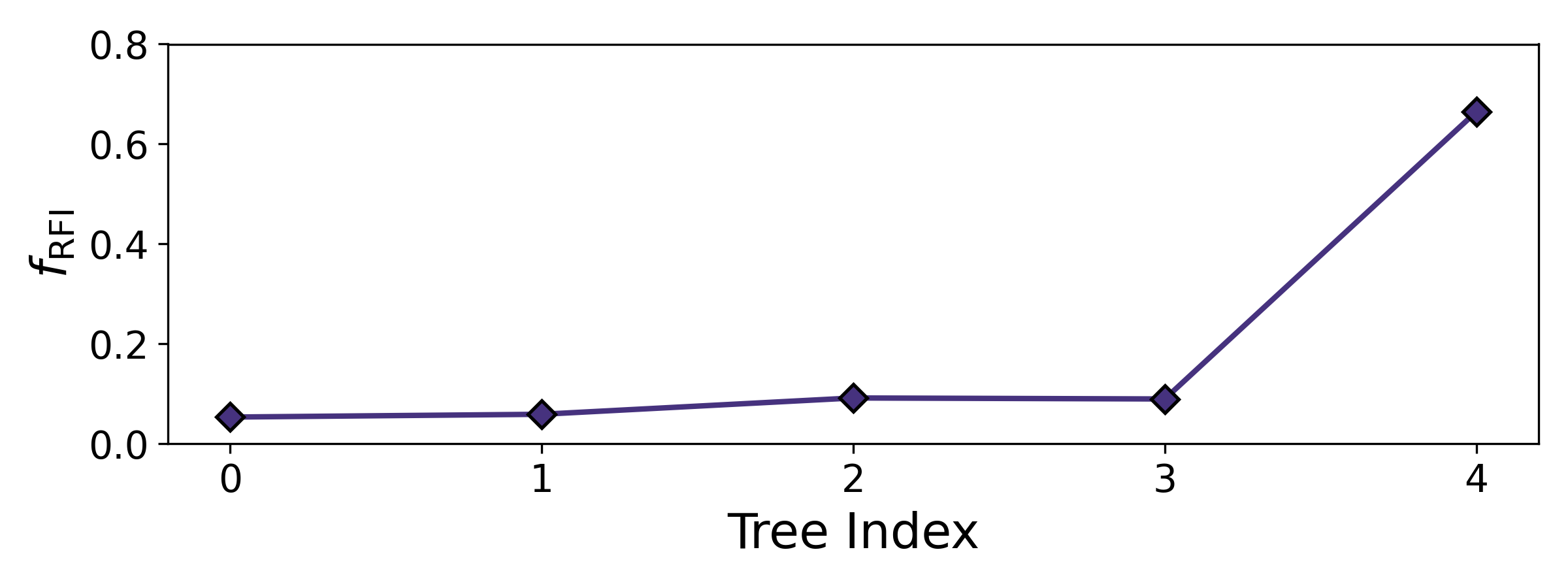}
    \caption{The fraction of events labelled as RFI ($f_\text{RFI}$) for each \texttt{bonsai} downsampling tree. While there is a slight increase in the fraction of events labelled as RFI going from trees 0--3, $f_\text{RFI}$ is very large ($\sim$75\%) for the largest downsampling tree.}
    \label{fig:rfi_v_tree}
\end{figure}


Figure \ref{fig:fluence_v_snr} compares the correlation of fluence and SNR for FRBs in the first CHIME/FRB catalog and the population of injections used for the inference analysis. To best compare the fluences of injections and catalog bursts, a cut was made on both samples to contain only pulses within the FWHM sensitivity of the most sensitive beam column. This cut was necessary because CHIME/FRB calibrates its FRBs using steady source transits from the center of beams in the most sensitive beam column only, making the derived fluences of bursts detected outside of these regions lower bounds.

The relationship between fluence and SNR should be consistent between injections and CHIME/FRB catalog bursts because the population inference analysis methods used in \cite{chime_catalog_1} employ the detection SNR as a proxy for the fluence. While it's difficult to statistically compare the 2D distributions, it is clear by eye that the two populations follow a similar trend.


\begin{figure}[t]
    \centering
    \includegraphics[width=\linewidth]{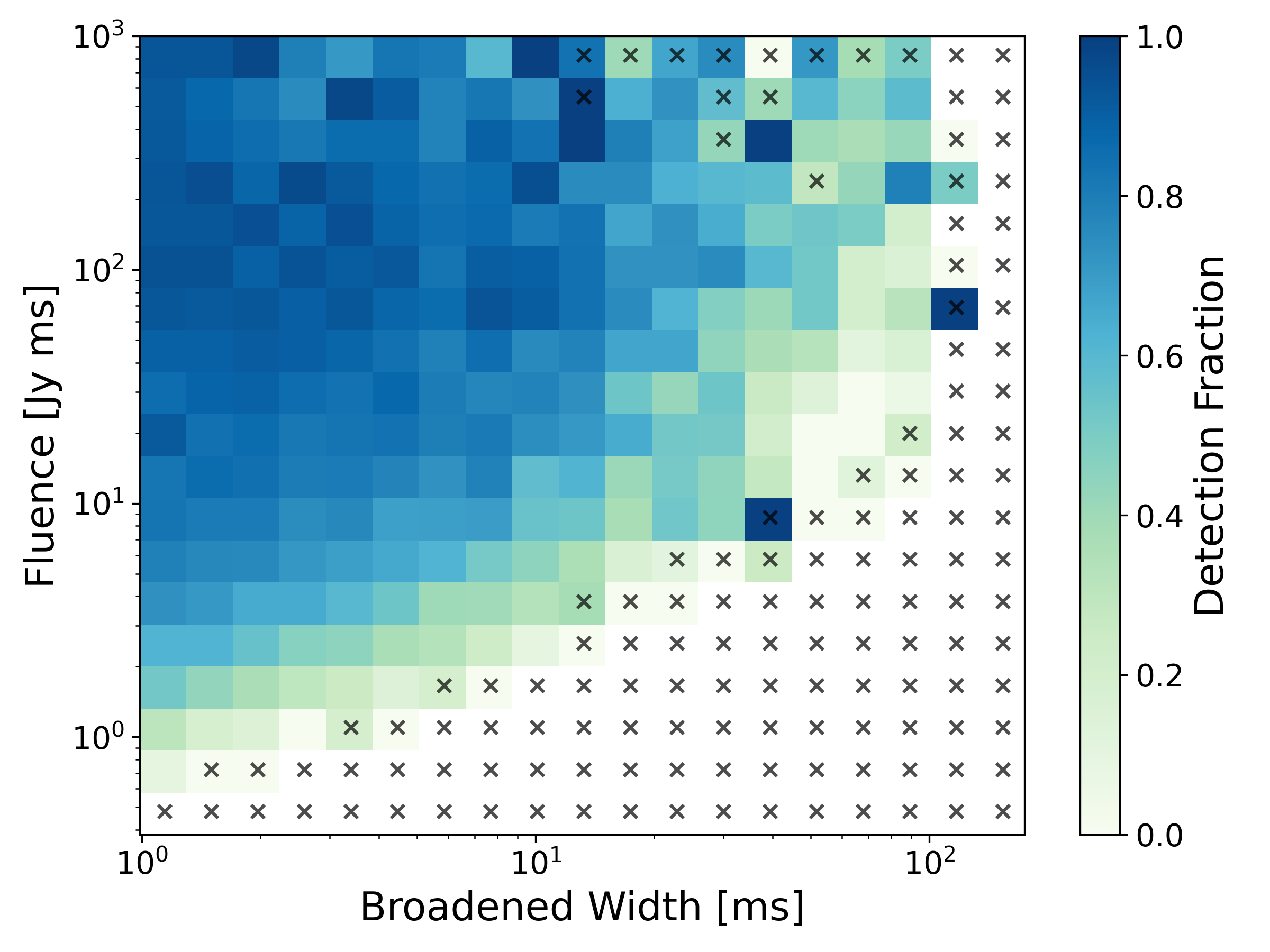}
    \caption{2D histogram showing the detection fraction of events as a function of fluence and broadened width. The histogram only shows bins with fluences $<10^3\ \text{Jy ms}$, as pulses with larger fluences made up only a small fraction of injections. Black crosses mark bins where there were fewer than 10 events and do not give a reliable detection fraction. For each fluence bin, there is an obvious gradient where detection fraction increases as pulse width decreases.}
    \label{fig:det_fraction_width}
\end{figure}

\begin{figure}[t]
    \centering
    \includegraphics[width=\linewidth]{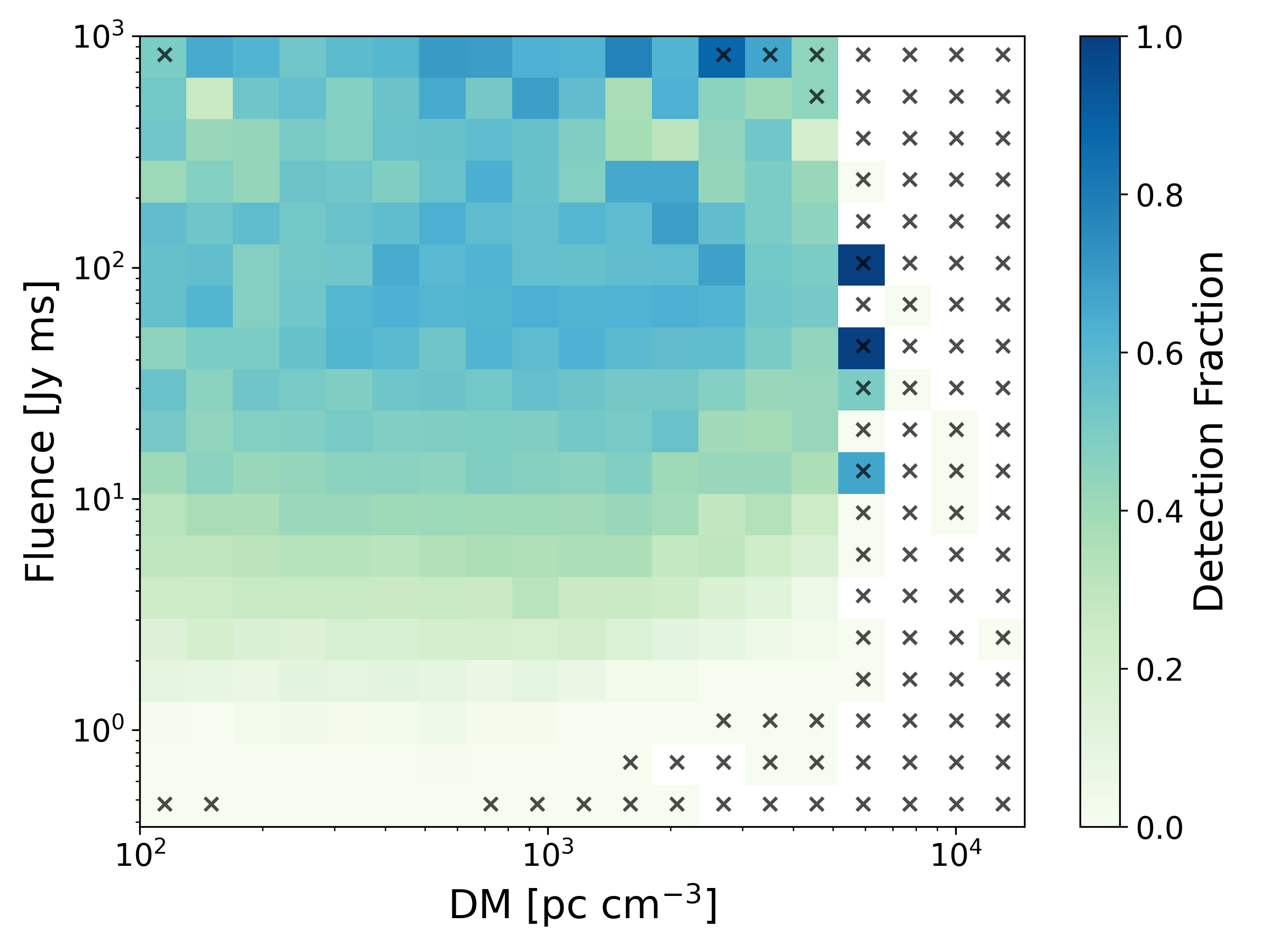}
    \caption{Similar to Figure \ref{fig:det_fraction_width}, but as a function of DM instead of broadened width. Only pulses with DM $>100\ \text{pc cm}^{-3}$ are shown, as few injections had a lower DM. There is no obvious correlation of detection fraction as a function of DM for any fluence bins, unlike in Figure \ref{fig:det_fraction_width}.}
    \label{fig:det_fraction_dm}
\end{figure}

\subsection{Pipeline Efficiency} \label{sec:efficiency}

Many aspects of the CHIME/FRB pipeline have intrinsic biases when detecting FRBs. This was noted first in \cite{abb+19a}, where many highly scattered events were detected even though simulations of the L1 pipeline indicated CHIME/FRB is biased against these kinds of events. However, the implementation of the injection system has allowed a greater understanding of where these biases arise throughout the CHIME/FRB pipeline, and how they impact the selection function described in \cite{chime_catalog_1}.

One way bright, low-DM FRBs can be missed by the pipeline is through clipping during L1 RFI mitigation. In particular, wide bursts at low-DM tend to have a significant amount of signal removed (proportional to the width and inversely proportional to the DM) when detrending along the frequency axis (Rafiei-Ravandi et al. 2022, in preparation). Though the algorithm has been iteratively improved, there is clearly a plateau in Figure \ref{fig:fluence_v_snr} past $\sim$100 Jy ms beyond which fluence is not a strong indicator of SNR. Though part of this plateau may be attributed to clipping, it may also be attributed to detections of bright events in the sidelobes of CHIME.

The most impactful bias of CHIME/FRB comes from classifying events with large broadened widths, particularly due to a high degree of scattering. We define the broadened width similar to \citep{ckj+17}:
\begin{equation} \label{eq:width_b}
    W_\text{b} = \sqrt{W_\text{i}^2 + t_\text{samp}^2 + t_\text{chan}^2 + \tau_{\text{600 MHz}}^2}\ .
\end{equation}
Here, $W_\text{i}$ is the intrinsic width, $t_\text{samp}$ is the sampling time, $t_\text{chan}$ is the dispersive delay within a frequency channel, and $\tau_{\text{600 MHz}}$ is the scattering time at 600 MHz. $W_\text{b}$ is used as a burst width metric because \texttt{bonsai} does not include a scattering tail in its search, making it difficult to disentangle the effects of scattering and width on pipeline sensitivity. Figure \ref{fig:rfi_v_tree} shows the fraction of injected pulses classified as RFI as a function of the \texttt{bonsai} tree index (see Table \ref{tab:bonsai_pars}). While the RFI fraction increases somewhat steadily from trees 0-3, its increase is dramatic in tree 4, where all of the widest pulses are detected. The selection effects as a function of width are also apparent in Figure \ref{fig:det_fraction_width}, where the detection fraction of events steadily decreases as width increases. The system's bias against the widest events can largely be explained by the data used to train L1b and L2's RFI excision algorithms. The astrophysical dataset for this training comprised primarily pulsar pulses detected by CHIME/FRB, which are typically narrow. By contrast, the RFI dataset contained many RFI signals resulting from aircraft transits, which are typically very wide. The result is a machine learning classifier that is strongly biased against highly scattered events, which is reflected in CHIME/FRB's selection function.

It is also interesting to consider any potential biases in the pipeline to high DM events. Events with large DMs are particularly useful for probing the IGM, and it is interesting to note that the first CHIME/FRB catalog only contains a few events with DM $>$ 1000 pc cm$^{-3}$. However, \cite{chime_catalog_1} did not find major selection effects as a function of DM. This is apparent in Figure \ref{fig:det_fraction_dm}, where the detection fraction does not depend strongly on DM. Since this is the case, it is an indication that FRBs with higher DMs than that detected by CHIME/FRB are fainter, as CHIME/FRB is not biased against them by their high DM alone. This will be explored in the future by including correlations between DM and brightness in the selection function \citep[see, e.g.,][]{jpm+21}.

\section{Summary \& Future Plans}


We have presented the injection system used in \cite{chime_catalog_1} to determine the selection function of the CHIME/FRB pipeline. The system was also used to monitor system sensitivity in the presence of factors such as a changing RFI environment and periodically re-trained RFI mitigation algorithms. We used the system to inject a population of $\sim$85,000 FRBs as described in \S\ref{sec:injection_architecture}. The synthetic pulses were calibrated in real time and drawn from populations based on the detected CHIME/FRB distributions.

We have learned what aspects of the CHIME/FRB pipeline impact the selection function:

\begin{itemize}
    \item The truncation of extreme values for very bright events (``clipping'') limits CHIME/FRB's sensitivity to the brightest FRBs. This effect, alongside bright events being detected in CHIME's low-sensitivity sidelobes, contribute to a plateau in the Fluence and SNR relationship seen in Figure \ref{fig:fluence_v_snr}.
    \item The CHIME/FRB pipeline has a bias against highly scattered events (Figures \ref{fig:rfi_v_tree} and \ref{fig:det_fraction_width}), likely resulting from the choice of data used to train the RFI mitigation algorithms. As described in \cite{chime_catalog_1}, this suggests a large population of highly scattered events remain presently undetected by CHIME/FRB.
    \item There is an absence of a strong selection against high DM events (Figure \ref{fig:det_fraction_dm}) suggesting that the dearth of high DM events in Catalog 1 is real and could arise from a DM-fluence correlation (Shin et al. 2022, in preparation).
\end{itemize}

In the future, we plan on expanding the injection system to operate on the entire array of CHIME/FRB beams at once, instead of only four replicated beams at a time. This will make the injection process far more efficient, allowing for a much larger sample of injections in preparation for the second CHIME/FRB catalog. With a larger sample, we will be able to generate an improved selection function, including correlations between important parameters such as DM and flux/scattering, and hence more accurately characterize the true population of FRBs based on the sample detected by CHIME/FRB. We also plan to use the information gained from the injection system to improve the CHIME/FRB pipeline's sensitivity. For example, the machine learning algorithms which currently bias the FRB search against detecting highly scattered bursts will be retrained on more robust astrophysical and RFI sources.

\acknowledgements

We acknowledge that CHIME is located on the traditional, ancestral, and unceded territory of the Syilx/Okanagan people. We are grateful to the staff of the Dominion Radio Astrophysical Observatory, which is operated by the National Research Council of Canada.  CHIME is funded by a grant from the Canada Foundation for Innovation (CFI) 2012 Leading Edge Fund (Project 31170) and by contributions from the provinces of British Columbia, Qu\'{e}bec and Ontario. The CHIME/FRB Project is funded by a grant from the CFI 2015 Innovation Fund (Project 33213) and by contributions from the provinces of British Columbia and Qu\'{e}bec, and by the Dunlap Institute for Astronomy and Astrophysics at the University of Toronto. Additional support was provided by the Canadian Institute for Advanced Research (CIFAR), McGill University and the McGill Space Institute thanks to the Trottier Family Foundation, and the University of British Columbia.

\allacks

\bibliographystyle{aasjournal}
\bibliography{frbrefs}

\end{document}